\documentclass[conference]{IEEEtran}
\IEEEoverridecommandlockouts
\usepackage{cite}
\usepackage{amsmath,amssymb,amsfonts}
\usepackage{array,multirow}
\usepackage{algorithmic}
\usepackage{graphicx}
\usepackage{textcomp}
\usepackage{xcolor}
\def\BibTeX{{\rm B\kern-.05em{\sc i\kern-.025em b}\kern-.08em
    T\kern-.1667em\lower.7ex\hbox{E}\kern-.125emX}}

\usepackage{listings}    

\usepackage{float}
\usepackage{syntax}
\usepackage{newfloat}
\usepackage{textcomp}
\usepackage{subfig}
\usepackage{tabularx}
\usepackage{balance}
\usepackage{url}

\usepackage{tikz}
\usepackage[hyperfootnotes=false]{hyperref} 

\newcommand\copyrighttext{
  \footnotesize \textcopyright 2022 IEEE. Personal use of this material is permitted.
  Permission from IEEE must be obtained for all other uses, in any current or future
  media, including reprinting/republishing this material for advertising or promotional
  purposes, creating new collective works, for resale or redistribution to servers or
  lists, or reuse of any copyrighted component of this work in other works.
  DOI: \href{https://doi.org/10.1109/IC2E55432.2022.00011}{https://doi.org/10.1109/IC2E55432.2022.00011}}
\newcommand\copyrightnotice{
\begin{tikzpicture}[remember picture,overlay]
\node[anchor=south,yshift=10pt] at (current page.south) {\fbox{\parbox{\dimexpr\textwidth-\fboxsep-\fboxrule\relax}{\copyrighttext}}};
\end{tikzpicture}
}
    
\begin{document}

\title{Towards Energy Consumption and Carbon Footprint Testing for AI-driven IoT Services
}

\author{\IEEEauthorblockN{ \parbox{\linewidth}{\centering  Demetris Trihinas\IEEEauthorrefmark{1}, Lauritz Thamsen\IEEEauthorrefmark{2}, Jossekin Beilharz\IEEEauthorrefmark{3}, and Moysis Symeonides\IEEEauthorrefmark{4}}}\\
\and
\IEEEauthorblockA{\IEEEauthorrefmark{1}
Dept. of Computer Science\\
University of Nicosia\\
trihinas.d@unic.ac.cy}
\and
\IEEEauthorblockA{\IEEEauthorrefmark{2}
School of Computing Science\\
University of Glasgow\\
lauritz.thamsen@glasgow.ac.uk}
\and
\IEEEauthorblockA{\IEEEauthorrefmark{3}
Hasso Plattner Institute\\
University of Potsdam\\
beilharz@acm.org}
\and
\IEEEauthorblockA{\IEEEauthorrefmark{4}
Dept. of Computer Science\\
University of Cyprus\\
msymeo03@ucy.ac.cy}

}

\maketitle
\copyrightnotice

\begin{abstract}
Energy consumption and carbon emissions are expected to be crucial factors for Internet of Things (IoT) applications. Both the scale and the geo-distribution keep increasing, while Artificial Intelligence (AI) further penetrates the “edge” in order to satisfy the need for highly-responsive and intelligent services. To date, several edge/fog emulators are catering for IoT testing by supporting the deployment and execution of AI-driven IoT services in consolidated test environments. These tools enable the configuration of infrastructures so that they closely resemble edge devices and IoT networks. However, energy consumption and carbon emissions estimations during the testing of AI services are still missing from the current state of IoT testing suites. This study highlights important questions that developers of AI-driven IoT services are in need of answers, along with a set of observations and challenges, aiming to help researchers designing IoT testing and benchmarking suites to cater to user needs.
\end{abstract}


\vspace*{-0.5\baselineskip}
\section{Introduction}
\label{sec:intro}
For a while now, IoT devices were considered sophisticated endpoints connecting the physical with the digital world, capable of serving data upstream to data centers. With recent advancements, however, IoT hardware is vastly improving, providing more compute power and storage capacity, while a plethora of devices is also embedding specialized accelerators now~\cite{Reuther2020}. This is moving the next generation of IoT services towards AI and transforming edge computing into Edge Intelligence~\cite{Deng2020}. %
However, large-scale AI is compute hungry. Since 2012, the amount of computational power used in the largest AI training is exponentially increasing, doubling every 4 months (compared to Moore’s Law 24-month doubling period).
Hence, even if IoT hardware is advancing, %
highly responsive AI is not a job for a single device. Therefore, it is no wonder that the scale and distribution of AI-driven IoT services are increasing.

Nonetheless, more compute effort results in more energy consumption and this may well result in more carbon emissions %
\cite{Henderson2020}\cite{Trihinas2021b}. Carbon emissions play a central role in climate change as they are directly responsible for the greenhouse effect~\cite{Lannelongue2021}. Already, data centres use an estimated 200TWh per year, equivalent to 1\% of the global energy demand~\cite{Jones2018}. Moreover, Google reports that approximately 15\% of it's energy use is attributed to AI/ML~\cite{Patterson2022}. Also, with Gartner indicating that 75\% of enterprise data are expected to be created and processed at the edge~\cite{Gartner}, one of the key challenges emerging is the migration %
to sustainable %
edge micro-DCs. However, the trend towards Edge Intelligence is not just difficult for developers, but might also have a significant negative impact on the environment~\cite{Georgiou2022}.

There exists a plethora of tools catering for the rapid and continuous testing of distributed IoT services~\cite{Beilharz2021}. These tools enable the seamless deployment of IoT services in consolidated environments where hosts can be configured to replicate heterogeneous edge devices and networks, while the service quality and fault tolerance can be evaluated at runtime through emulation.
Still, with climate change initiatives being adopted by ICT organisations (e.g., CarbonTrust standard~\cite{carbon-trust}) and with recent events (e.g., Ukraine war) intensifying the move towards carbon-neutral commitments (e.g., EU green deal~\cite{green-deal}, UK net-zero~\cite{net-zero}), low carbon emissions will be an important requirement for AI-driven IoT services. Yet, IoT testing tools do not cater for in-depth benchmarking of energy consumption and carbon emissions, thus excluding the environmental footprint from the testing~\cite{Beilharz2021}. Part of the reason testing tools do not report energy and carbon emissions are the complexities in calculating accurate estimates. These footprints require an understanding of emissions from energy grids, assessing power drawn for computation and communication, as well as navigating and integrating multiple different tools~\cite{Henderson2020}. This leaves users puzzled in regards to the energy consumption and carbon footprint of their AI-driven IoT services.

This paper highlights the need and challenges that come with deploying AI-driven IoT services in edge computing settings in terms of energy consumption and carbon footprints. For this, we present a background on the metrics required to give energy and carbon estimations. Next, we introduce an edge-driven object detection application that serves as a reference point for the following sections. Then, we discuss important questions that IoT service developers are in need of answers to and outline central challenges that should be addressed by the next generation of edge testing tools. 

\section{Background}
\label{sec:background}
The following provides an overview of the background knowledge to apprehend energy and carbon emission estimation. With this overview, we show that there are a variety of factors to consider so that IoT testing tools can provide accurate estimations. %

\vspace*{-0.25\baselineskip}
\subsection{Energy Consumption}
\label{sec:energy}
Energy consumption, denoted as $E$ and measured in \textit{Joules}, is defined as the amount of energy required by a computing system to execute a specific task. Energy is calculated with:
\begin{equation}
    E = P \cdot t
\end{equation}

In this, $P$, measured in \textit{Watts}, is denoted as the power drawn by the computing system and $t$, measured in \textit{seconds}, is the total amount of time required by the system to finish the desired task. In line with the above definition, $E$ can also be measured in $Wh$, denoted as \textit{Watt-hours}.

Power usage is reported as the sum of $P_{idle} + P_{dyn}$, where $P_{idle}$ denotes the load independent power drawn by the computing system, even if no task is under execution, and $P_{dyn}$ is load dependent. %
Assuming the task is a software service (i.e., ML training) the key components contributing to $P_{dyn}$ are the use of processors, memory and graphic accelerators. The latter, when available, consume the overwhelming majority of the power drawn from the energy source~\cite{lenovo-bigdata}. %
To account for other components (i.e., cooling) in a (micro) datacenter setting, one can rely on the Power Usage Effectiveness (PUE) factor~\cite{Strubell2019}. The PUE is an industry standard defined as the total energy needed for all aspects of operation, including cooling, divided by the energy used directly for computing:
\begin{equation}
    PUE = \frac{Total\:Energy\:Consumption}{Computing\:Task\:Consumption}
\end{equation}

This factor scales the available power metrics by a mean projected overhead for related power consumption. Therefore, energy consumed by a computing system can be extended to:
\begin{equation}
    E = PUE\int_{0}^{t}(P_{idle}+P_{dyn})\:dt
\end{equation}

With a PUE closer to 1.0, energy is consumed purely for the desired purpose. The mean PUE for data centres in 2020 was 1.58~\cite{Ascierto2018}, with cloud providers, i.e., Google and AWS, reporting more efficient values in the range 1.1-1.2~\cite{Patterson2022}. %
However, current edge micro-DCs are not tailored for energy-efficiency, e.g., due to their high degree of heterogeneity and inefficient cooling, with reports indicating a mean PUE of 2.0 for 2021 and projections for 2040 lowering this to just 1.5~\cite{Schneider}. Therefore, \textit{in-place data processing and AI will face the challenge to optimize energy consumption at the edge}.

\subsection{Carbon Emissions}
\label{sec:carbon}

\begin{figure}[t]
    \centering
    \includegraphics[width=0.75\linewidth]{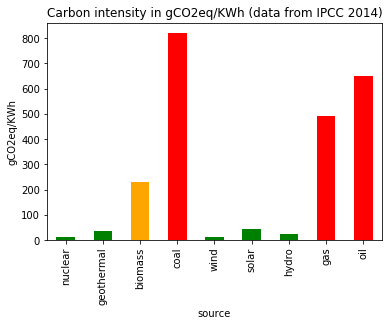}
    \caption{Carbon intensity of various energy sources}
    \label{fig:ci-sources}
    \vspace*{-1\baselineskip}
\end{figure}

Carbon emissions are characterized by the \textit{Carbon Intensity} coefficient ($CI$). The $CI$ is measured in $gCO_2eq/kWh$ and accounts for the \textit{grams} of carbon dioxide equivalent greenhouse gas emissions ($CO_2eq$) released for every $KWh$ of consumed energy. %
Carbon intensity is considered a standardized measure describing the ``cleanliness" of the energy consumed by a system (i.e., edge micro-DC)~\cite{Qiu2020}. Hence, $CO_2eq$ emissions are the product of two factors: They can be calculated by multiplying the amount of energy consumed, expressed in KWh, with the CI coefficient.
\begin{equation}
    CO_2eq = E\cdot CI
\end{equation}

Still, the CI coefficient can be hard to obtain as it depends on the sources the energy grid draws power from. In 2014, the Intergovernmental Panel on Climate Change (IPCC) harmonized the carbon intensity of the key electricity generating sources with these reference values shown in Figure~\ref{fig:ci-sources}. Many national and regional authorities report a fixed carbon intensity coefficient, even for the duration of an entire year. This coefficient is calculated after aggregating and weighting the region's energy production ($\vec{\beta}$) from various sources ($c\in S$), often denoted as the \textit{energy mix}. 
\begin{equation}
    CI_{grid} = \sum_j^{|S|}\beta_j\cdot c_j
\end{equation}

For example, the latest report by the European Environment Agency states that for 2020 the $CO_2$ emission intensity for Cyprus was 621, for Germany 311, and for Sweden 13 $gCO_2/KWh$, respectively~\cite{eea}. While the CI coefficient can greatly differ across countries and regions, as we will show, carbon emissions can vary even throughout the day. These \textit{large temporal variations in carbon intensity, due to changes in the energy mix, make it difficult to acquire and calculate accurate estimations during software testing to reduce and optimize carbon emissions}.

\section{Research Questions \& Observations}
\label{sec:observations}
This section highlights key questions that users are faced with when assessing the energy consumption and carbon emissions of AI-driven IoT services during application testing, after introducing an edge-AI reference scenario.

\subsection{Reference Scenario}
\label{sec:scenario}

As a reference scenario, let us consider an AI-driven IoT service where numerous geo-dispersed IoTs (i.e., cameras, drones) can be employed for object detection at a city-scale level. To aid recurrent model training at a neighborhood level, improve privacy and facilitate local device coordination, edge micro-DCs are deployed and scattered across the city. 

Unless otherwise stated, we will assume a baseline configuration where the adopted edge micro-DC is powered by a DELL PowerEdge R610 server (12cores@2.4GHz, 12GB memory, max 330W) and equipped with a Nvidia T4 GPU (320 tensor cores, 16GB, max 70W). %
For the ML task we employ the TensorFlow benchmark suite to output a CNN model for object detection trained with the ImageNet dataset (144GB, 1.3M images)~\cite{tf-bench}. In turn, TensorFlow Lite is used to deploy the trained models %
for on-device inference.

\subsection{When to train a model?}
\label{sec:when}
\begin{figure}[t]
   \begin{minipage}{0.5\textwidth}
        \centering
        \includegraphics[width=.65\textwidth, trim=0 18 0 0]{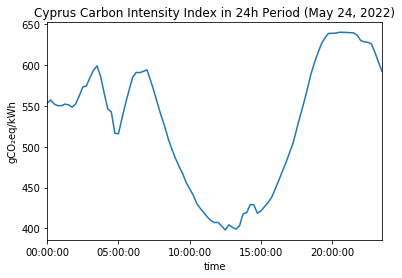}
        \caption{Cyprus 24h carbon intensity}
        \label{fig:cy-ci}
    \end{minipage}
   \begin{minipage}{0.5\textwidth}
        \centering
        \includegraphics[width=.65\textwidth, trim=0 18 0 0]{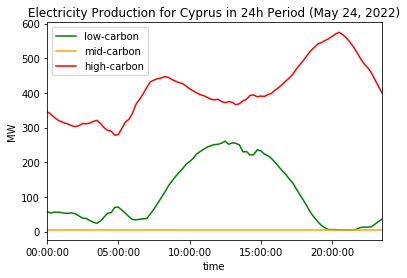}
        \caption{Cyprus 24h electricity production}
        \label{fig:cy-eprod}
   \end{minipage}
   \vspace*{-1.\baselineskip}
\end{figure}

Many national and regional authorities report a fixed carbon intensity coefficient, even for the duration of an entire year. %
However, carbon intensity may drastically fluctuate even throughout various periods of the day. Figure~\ref{fig:cy-ci} depicts an example, where the carbon intensity is computed for a given day in Cyprus. The reason carbon intensity fluctuates is that it depends on energy production and specifically the sources powering the energy grid. Figure~\ref{fig:cy-eprod} illustrates this for Cyprus, where the use of low-carbon energy sources (solar and wind) start and peak during day time with the majority originating from solar power ($>$80\%). As such, the amount of carbon emissions attributed to model training can drastically differ, depending on when a model is trained. 

\begin{figure}[t]
    \centering
    \includegraphics[width=0.75\linewidth, trim=0 18 0 0]{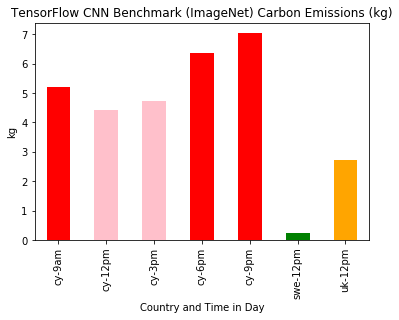}
    \caption{Carbon footprint for different countries and time in day}
    \label{fig:carbon-kg}
    \vspace*{-1.25\baselineskip}
\end{figure}

With the above in mind, let us consider power measurements extracted from training the ML model of the reference scenario for almost 4.5 hours on the described edge micro-DC. Figure~\ref{fig:carbon-kg} depicts the estimated carbon footprint of the ML training for different periods of the (same) day in Cyprus. From this figure we observe that training the model during midday features a carbon footprint that is 1.64kg less that initiating training at 6pm and 2.3kg less than at 9pm.

\textbf{Observation}: When a model is trained can have a significant impact in terms of carbon emissions. The footprint depends on the energy sources currently in use by the grid that power is drawn from. Therefore, organisations wanting to reduce their carbon footprint should consider training their ML models when low-carbon energy sources are producing power for the energy grid that the computing system consumes energy from.

\subsection{Where to train a model?}
\label{sec:where}
\begin{figure}[t]
   \begin{minipage}{0.5\textwidth}
        \centering
        \includegraphics[width=.65\textwidth, trim=0 18 0 0]{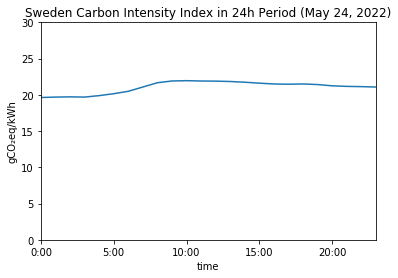}
        \caption{Sweden 24h carbon intensity}
        \label{fig:swe-ci}
    \end{minipage}
   \begin{minipage}{0.5\textwidth}
        \centering
        \includegraphics[width=.65\textwidth, trim=0 18 0 0]{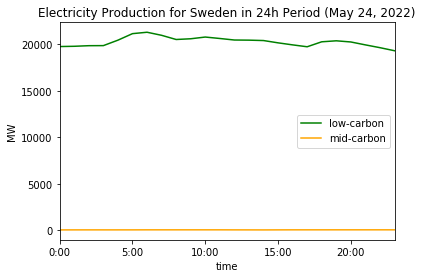}
        \caption{Sweden 24h electricity production}
        \label{fig:swe-eprod}
   \end{minipage}
   \vspace*{-1.5\baselineskip}
\end{figure}

At this point, one may advocate that Cyprus may not be an ideal country to train ML models since it makes heavy use of high-carbon energy sources (i.e., oil)~\cite{tsoc}. Therefore,  other European countries could be explored. %
As an example, Figures~\ref{fig:swe-ci} and \ref{fig:swe-eprod} depict the energy production and carbon intensity for Sweden during the same referenced day. Sweden is the EU member state with the lowest carbon footprint~\cite{eea}. From these figures we observe that 99\% of Sweden's energy production comes from low-carbon sources (approx. 34\% hydro, 45\% nuclear and the rest is a mix of wind, geothermal and solar energy). Sweden's carbon intensity coefficient is also relatively stable throughout the day. This is ideal in the cases model training simply cannot wait. Based on these estimates, the same CNN model will be trained in Sweden with a carbon footprint of just 0.21-0.24kg, no matter the time throughout the day. This footprint is equivalent to a 95\% reduction to ML training, even, midday in Cyprus. 

\textbf{Observation}: The ML model training process can have a significantly different environmental footprint over different locations. Therefore, organisations wanting to reduce the environmental impact of their model training should explore potential advantages of moving their workloads to low-carbon grid energy.

\subsection{What is the impact of model inference?}
\label{sec:models}
The vast majority of AI studies consider accuracy as the pertinent performance measure, ignoring the impact ML has on deployed devices and the environment~\cite{GreenAI}. Towards this, let us consider three state-of-the-art model architectures for ML classification that can be used for object detection. Specifically, we will consider ResNet50 (used in Sections~\ref{sec:when} and \ref{sec:where}), SqueezeNet and MobileNetV2. 

We now focus on ML inference and its impact to the reference edge micro-DC. For this, we use binary builds from the TensorFlow model zoo for the adopted architectures trained using the reference dataset~\cite{tf-model-zoo}.  Table~\ref{table:models} depicts the results. We immediately observe the high accuracy of ResNet50, but through the additional metrics we can also label it as both energy-hungry and latent in terms of inference. In turn, SqueezeNet is a CNN architecture designed for model compactness, delivering a low memory footprint suitable even for small IoT settings. In line with this, inference time is less than a fifth compared to ResNet50. However, this performance comes with a drop in accuracy of more than 20\% compared to ResNet50. In contrast, MobileNetV2 presents an architecture that achieves a small runtime footprint, with only a 3\% accuracy reduction, and when compared to ResNet50 reduces energy consumption by almost 90\%.

\textbf{Observation}: There is often more than one ML model architecture that can be used for a task but the impact on system utilization as well as the energy footprint (and consequently battery autonomy) can differ significantly. We must move from solely looking at accuracy metrics to also examining resource overhead, energy footprint and environmental impact metrics when testing and releasing models.

\begin{table*}[t]
\centering
\caption{Performance comparison of various ML model architectures during inference}
\label{table:models}
\begin{tabular}{lccccc}
\hline
\multicolumn{1}{c}{\textbf{Model}} & \textbf{\begin{tabular}[c]{@{}c@{}}Params\\ (Million)\end{tabular}} & \textbf{\begin{tabular}[c]{@{}c@{}}Size\\ (MB)\end{tabular}} & \textbf{\begin{tabular}[c]{@{}c@{}}Accuracy\\ (\%)\end{tabular}} & \textbf{\begin{tabular}[c]{@{}c@{}}Mean Inference Time\\ (ms)\end{tabular}} & \textbf{\begin{tabular}[c]{@{}c@{}}Energy Consumption\\ (Wh per 5K images)\end{tabular}} \\ \hline
ResNet50                           & 25.6                                                                & 99                                                           & 73                                                               & 1116                                                                              & 373                                                                                      \\
SqueezeNet                         & 1.25                                                                & 5                                                            & 52                                                               & 212                                                                               & 51                                                                                       \\
MobileNetV2                        & 3.4                                                                 & 14                                                           & 70                                                               & 143                                                                               & 39                                                                                       \\ \hline
\end{tabular}
\vspace*{-1\baselineskip}
\end{table*}

\subsection{How does resource heterogeneity impact ML inference?}
Resource heterogeneity in IoT is to be expected in many different forms (i.e., resource availability, different devices, high-priority base loads). After conducting the Section~\ref{sec:models} experiments that examine both accuracy and system measures, the MobileNetV2 architecture seems to come off favorably, at least for the reference scenario. In the next line of experimentation we employ MobileNetV2 and study the impact of resource heterogeneity on inference and power consumption. For this, we consider both different devices and computational power by capping the CPUs available. We did this on the server of our reference scenario and on a Raspberry Pi 4 model B. The open-source Fogify framework~\cite{Symeonides2020} was used for this as it supports resource capping on host environments and provides a plugin interface so that custom monitoring can be implemented easily~\cite{trihinas2016}. These features were used to integrate smart energy meters as Fogify does not currently support energy measurements out-of-the-box. Table~\ref{table:devices} depicts our results. After studying the table, at first, one can observe that for both devices as compute availability increases, inference time is reduced, while the power consumption increases. Yet, the behavior of both metrics cannot easily be characterized with a mathematical distribution. Several hardware phenomena actually take place under the hood (i.e., dynamic voltage scaling, cooling, etc.) which manifest in trade-offs that must be examined carefully. Moreover, taking a quick glance at both devices, we see that from a sustainable computing perspective, employing a Raspberry Pi 4 at full capacity can be much better than using a server at 50\% capacity with power reduced by 95\% for a 30\% impact on inference time (60ms).

\textbf{Observation}: Different devices and also different resource configurations can have a considerable impact on both a model's inference performance as well as power required for computations. These trade-offs should be tested to find the configurations that deliver the required performance and reliability at the lowest environmental impact.

\begin{table}[t]
\centering
\caption{Compute availability impact on ML inference}
\label{table:devices}
\begin{tabular}{cccc}
\hline
\textbf{Device} & \textbf{Cores} & \textbf{\begin{tabular}[c]{@{}c@{}}Mean Inference Time\\ (ms)\end{tabular}} & \textbf{\begin{tabular}[c]{@{}c@{}}Mean Power Drawn\\ (W)\end{tabular}} \\ \hline
Server          & 3              & 432                                                                               & 87                                                                      \\
                & 6              & 210                                                                               & 91                                                                      \\
                & 9              & 152                                                                               & 139                                                                     \\
                & 12             & 143                                                                               & 197                                                                     \\ \hline
RPi             & 1              & 693                                                                               & 4.2                                                                     \\
                & 2              & 331                                                                               & 5.1                                                                     \\
                & 3              & 273                                                                               & 5.2                                                                     \\
                & 4              & 269                                                                               & 5.9                                                                     \\ \hline
\end{tabular}
\vspace*{-1.25\baselineskip}
\end{table}

\vspace*{-0.25\baselineskip}
\section{Challenges for IoT Testing Frameworks}

We imagine IoT testing tools will support the assessment of the energy consumption and carbon emissions for future AI-driven IoT applications. For this, testing tools will likely monitor the resource usage of applications running on emulated infrastructures.
To then translate resource usage into energy consumption, the testing tools will presumably use power models. To translate energy into carbon emissions, these tools will need further data such as the carbon intensity of the energy mix during task execution. While this idea of performance testing of actual software for estimating carbon emissions is straightforward, there are many challenges in integrating this into IoT testing tools so that users get answers to the previously raised questions. This section lists and discusses challenges we foresee for effectively integrating energy consumption and carbon emissions testing into the current state of IoT testing tools. These challenges can be a starting point for new research in the area of IoT testing. %
However, this list should not be considered final or complete.
For instance, emissions can also be associated with the production of devices that cannot be captured by simply translating current resource usage to power consumption and emissions.%

\subsection{Configuration of Power Models}

If power models are used to translate resource usage into energy consumed, these models will need to be supplied to IoT testing tools. Configuring appropriate power models can quickly become a larger configuration effort for users of IoT testing tools. This is especially true when more resources than individual devices are to be assessed for AI services. Imagine large-scale heterogeneous IoT deployments that span different devices, edge and cloud resources, graphical AI  accelerators, as well as a variety of local and wide-area networking links. Furthermore, this task becomes even more complex when infrastructure is distributed across geographic regions.

Data points to model the power consumption of particular %
resources can often be found online and in the literature. The community could also benefit by sharing power models for common infrastructure components in repositories. However, the majority of these approaches neglect the power required for housing resources. For instance, considerable energy is also required to cool compute resources under load~\cite{Gao2014}. Another way to get accurate models is to measure energy consumption on particular resources, extracting load-dependent power metrics of the usage of CPUs, GPUs, memory, and other resources. A study from Google found that power/carbon calculators overestimate calculations by not updating model parameters (i.e., CI, PUE)~\cite{Patterson2021}. However, obtaining measurements across large, geo-distributed IoT deployments is a costly endeavor.

\textbf{Recommendation}: IoT testing tools should support their users in finding power models for IoT infrastructures, leveraging for instance shared repositories for common infrastructure components, while also enabling the integration of real measurements to fine-tune power model accuracy.%

\subsection{Integration of Carbon Emissions Data}

If the carbon intensity of an energy mix is used to translate energy consumed into emissions, then this data must be integrated into IoT testing tools. Some energy providers now release production data and some even have APIs (i.e., data for Cyprus features a 15min granularity), while there also commercial online services that aggregate carbon data for many regions of the world~\cite{electricitymaps}. However, there is no data available for all providers or regions, while schemes for extraction and usage differ. Furthermore, this data can be historical when testing a range of scenarios based on past compositions of energy mixes, or, can be a forecast, when testing for instance scenarios based on the availability of renewable energy within the next day. Similar to power models, this will need to be configured with testing results dependent on these configurations.
Moreover, infrastructures might be supplied by multiple power sources. Computing infrastructure could, for example, be powered by on-site renewable energy sources such as wind or solar, while also being connected to a public energy grid. This would make it even harder to configure where the energy for computational and communication resources is coming from at any given time and, thus, how much emissions are associated with any energy consumption.

Another challenge with using carbon intensity to convert energy consumption into emissions is that this coefficient only captures the amount of emissions of the entire energy mix: Based on the energy mix each KWh used in the system is associated with the \emph{same} amount of $CO_2eq$ greenhouse gases. However, an energy mix does not really specify which energy is used for any particular consumption. It is fully possible that we might explore shifting larger workloads to a time or a place with a low-carbon energy mix, believing this will save emissions, yet in reality for this additional energy consumption, we could end up having more power generated from, for instance, fossil fuels like coal or gas. This is a known issue with the metric of carbon intensity, but since it is much harder to estimate where any particular energy is coming from, carbon-aware computing approaches still resort to simply using carbon intensity as their signal~\cite{GoogleCarbonAwareComputing,WorkloadShiftingCarbonEmissions,acun2022carbon}. 

\textbf{Recommendation}: IoT testing tools should provide a ``quick-start" carbon estimation process to quickly get novice users results, yet also convey limitations of these calculations. In turn, they should also support the integration of more representative carbon intensity data for changing energy mixes.

\subsection{Testing Trade-Offs}

There are many significant trade-offs that users must develop an understanding for testing AI-driven IoT applications. One example is the trade-off to save energy by reducing the frequency of model re-training, as well as moving model training to times when low-carbon energy is available (observation~1). %
However, waiting for the ideal moment to train a model presents itself with challenges. The purpose of retraining a model is %
so that changes in the data distribution, also known as concept drift, are timely reflected in the model during inference. Hence, finding the sweet spot between accuracy and energy saving boils down to how long training can be postponed without hampering accuracy.

Another significant trade-off to investigate is whether an AI application should be moved to a location with a better availability of low-carbon energy (observation 2). However, there are costs in terms of time, energy, and emissions for migration~\cite{Trihinas2017}. Often, AI requires large volumes of training data, the models can have a significant size, and moving data can incur delays. For example, the ImageNet dataset is 144GB and the time to move it over a 100Mbps link is 3.5 hours. Assessing this trade-off can be difficult as the costs of moving data amortize only over time, but energy systems are often highly dynamic, so it is not clear whether anticipated benefits actually accrue. In turn, moving data across regions is not a simple
process with potential legal and
privacy aspects contradicting the benefits of in-place processing and edge intelligence in general. Finally, another trade-off comes by opting for the use of a lightweight model drawing considerable energy savings (observation 3). However, saving energy must never come at the cost of significantly degrading accuracy. %

\textbf{Recommendation}: IoT testing tools should aid in the assessment of trade-offs so that users can take informative decisions. This includes finding trade-off sweet spots during testing to avoid catastrophic results in production such as significant overheads or accuracy degradation. To do so, testing tools must integrate, and maybe even innovate, a variety of benchmark measures so that users can sufficiently evaluate, beyond model accuracy, the impact of a model on system utilization, energy consumption and carbon footprint.

\vspace*{-0.25\baselineskip}
\subsection{Selection and Execution of Test Cases}
The energy footprint and carbon emissions of AI applications that run on large IoT infrastructures, powered by particular energy systems, depend on the particular applications, infrastructures, and energy mix. Moreover, all three factors are highly dynamic. Therefore, test results will rightly be different depending on the time, date, and location. Workloads, mobile infrastructures, and energy mixes change in seasonal patterns, while there are also trends that lead to more permanent change. This can include increasing usage of an application over time or a shift towards more renewable energy sources in a regional energy system. This makes it important to configure sets of test cases that cover different possible situations.

At the same time, testing the footprint of an AI-driven IoT application will come with its own footprint, especially when considering multiple scenarios on large-scale emulated infrastructures. There are strong benefits to testing software continuously, with tests triggered on change of application code or infrastructure definitions. Yet, if the goal is to develop and operate AI services more resource-efficient and sustainable, then testing towards this objective should not consume large amounts of energy with considerable additional emissions.

\textbf{Recommendation}: IoT testing tools should help developers select a good set of representative scenarios over possible times and locations, balancing the requirement to cover the different possible situations with the goal to use few resources to assess these scenarios. If indeed multiple variants of a basic scenario are tested, tools should further provide feedback that effectively generalizes from any of the specific situations.

\section{Related Work}

The related work that could help to evaluate the energy consumption and carbon footprint of AI-driven IoT services falls into three categories. 

\emph{Fog and edge emulators}, like Fogify~\cite{Symeonides2020}, Mockfog~\cite{hasenburg2019mockfog}, Marvis~\cite{beilharz2021towards} or IOTier~\cite{nikolaidis2021iotier}, enable the testing of actual IoT software behaviour in different scenarios by shaping the infrastructure to replicate heterogeneous edge devices and networks. 
While few fog emulators include energy consumption estimation~\cite{Beilharz2021}, there is currently no tool that addresses the particular challenges of estimating energy consumption and carbon emissions of AI-driven IoT services. 

\emph{Fog and edge simulators} mostly focus on the behavioral simulation of computational and networking resources and many, like EdgeCloudSim~\cite{sonmez2018edgecloudsim} or IOTSim~\cite{zeng2017}, do not support energy and emission modeling.
Others, like iFogSim~\cite{gupta2017ifogsim} and PureEdgeSim~\cite{mechalikh2019pureedgesim}, can simulate power utilization, but only for system entities such as cloud and fog nodes (iFogSim) or edge devices and their local network (PureEdgeSim).
On the other hand, LEAF~\cite{wiesner2021leaf} is a simulator that primarily focuses on the simulation of power consumption of complete edge and fog systems.
While such fog and edge simulators can provide valuable insights by allowing specific scenarios to be assessed, modeling actual software using these simulators is often difficult and requires careful configuration of simulation parameters. 
Furthermore, there are no simulators that focus on edge AI use cases or provide carbon emission estimations. 
 
Finally, we note that there are \emph{carbon emission estimation tools} that attempt to estimate the emissions based high-level job characteristics. 
Some of these estimation tools include carbon intensity data from cloud providers~\cite{TeadsCarbonEstimator}, that allow for the emissions estimation of compute nodes based on the type of cloud instance, duration of usage and location of the datacenter. 
Others~\cite{Lacoste2019} specialize on ML applications using the energy consumption of popular GPUs as basis. 
Carbontracker~\cite{anthony2020carbontracker} measures the power consumption of one or a few training epochs to predict the total power usage of NN-based applications. 
All these emission estimation tools focus on the power consumption of computing on a few central compute nodes, instead of distributed edge AI workloads.  

\section{Summary}
This work presents a study of the IoT testing landscape for energy- and carbon-aware AI-driven IoT services. With the increasing adoption of compute-hungry AI into IoT services, we need novel techniques that balance energy and compute demands to achieve more resource-efficient and sustainable IoT services. Towards this, we highlighted key questions for developing energy-efficient and carbon-aware AI-driven IoT services. After introducing these questions with concrete examples,  we present our observations and elaborate on what we believe must be the future directions and objectives that should be addressed with the next generation of EdgeAI emulators and IoT testing tools. In the future, we want to address these directions with new methods that can be integrated by open and popular testing suites.

\smallskip
\smallskip
\small{\noindent \textbf{Acknowledgement.} This work is partially supported by the University of Nicosia Seed Grant Scheme for the FlockAI project.}

\bibliographystyle{IEEEtran}
\bibliography{refs.bib}

\begin{thebibliography}{10}
\providecommand{\url}[1]{#1}
\csname url@samestyle\endcsname
\providecommand{\newblock}{\relax}
\providecommand{\bibinfo}[2]{#2}
\providecommand{\BIBentrySTDinterwordspacing}{\spaceskip=0pt\relax}
\providecommand{\BIBentryALTinterwordstretchfactor}{4}
\providecommand{\BIBentryALTinterwordspacing}{\spaceskip=\fontdimen2\font plus
\BIBentryALTinterwordstretchfactor\fontdimen3\font minus
  \fontdimen4\font\relax}
\providecommand{\BIBforeignlanguage}[2]{{%
\expandafter\ifx\csname l@#1\endcsname\relax
\typeout{** WARNING: IEEEtran.bst: No hyphenation pattern has been}%
\typeout{** loaded for the language `#1'. Using the pattern for}%
\typeout{** the default language instead.}%
\else
\language=\csname l@#1\endcsname
\fi
#2}}
\providecommand{\BIBdecl}{\relax}
\BIBdecl

\bibitem{Reuther2020}
A.~Reuther, P.~Michaleas, M.~Jones, V.~Gadepally, S.~Samsi, and J.~Kepner,
  ``Survey of machine learning accelerators,'' in \emph{2020 IEEE high
  performance extreme computing conference (HPEC)}.\hskip 1em plus 0.5em minus
  0.4em\relax IEEE, 2020.

\bibitem{Deng2020}
S.~Deng, H.~Zhao, W.~Fang, J.~Yin, S.~Dustdar, and A.~Y. Zomaya, ``Edge
  intelligence: The confluence of edge computing and artificial intelligence,''
  \emph{IEEE Internet of Things Journal}, vol.~7, no.~8, 2020.

\bibitem{Henderson2020}
P.~Henderson, J.~Hu, J.~Romoff, E.~Brunskill, D.~Jurafsky, and J.~Pineau,
  ``Towards the systematic reporting of the energy and carbon footprints of
  machine learning,'' \emph{Journal of Machine Learning Research}, vol.~21, no.
  248, 2020.

\bibitem{Trihinas2021b}
D.~Trihinas, M.~Agathocleous, K.~Avogian, and I.~Katakis, ``{FlockAI: A Testing
  Suite for ML-Driven Drone Applications},'' \emph{Future Internet}, vol.~13,
  no.~12, 2021.

\bibitem{Lannelongue2021}
L.~Lannelongue, J.~Grealey, and M.~Inouye, ``Green algorithms: quantifying the
  carbon footprint of computation,'' \emph{Advanced science}, vol.~8, no.~12,
  2021.

\bibitem{Jones2018}
N.~Jones, ``How to stop data centres from gobbling up the world’s
  electricity,'' \emph{Nature}, vol. 561, 2018.

\bibitem{Patterson2022}
D.~Patterson, J.~Gonzalez, U.~Hölzle, Q.~H. Le, C.~Liang, L.-M. Munguia,
  D.~Rothchild, D.~So, M.~Texier, and J.~Dean, ``{The Carbon Footprint of
  Machine Learning Training Will Plateau, Then Shrink},'' 4 2022.

\bibitem{Gartner}
\BIBentryALTinterwordspacing
Gartner, ``{What Edge Computing Means for Infrastructure and Operations
  Leaders},'' 2018. [Online]. Available:
  \url{https://www.gartner.com/smarterwithgartner/what-edge-computing-means-for-infrastructure-and-operations-leaders}
\BIBentrySTDinterwordspacing

\bibitem{Georgiou2022}
S.~Georgiou, M.~Kechagia, T.~Sharma, F.~Sarro, and Y.~Zou, ``Green ai: Do deep
  learning frameworks have different costs?''\hskip 1em plus 0.5em minus
  0.4em\relax ACM: Association for Computing Machinery, 2022.

\bibitem{Beilharz2021}
J.~Beilharz, P.~Wiesner, A.~Boockmeyer, L.~Pirl, D.~Friedenberger,
  F.~Brokhausen, I.~Behnke, A.~Polze, and L.~Thamsen, ``Continuously testing
  distributed iot systems: An overview of the state of the art,'' in \emph{19th
  International Conference on Service-Oriented Computing (ICSOC)
  Workshops}.\hskip 1em plus 0.5em minus 0.4em\relax Springer, 2022, p. to
  appear.

\bibitem{carbon-trust}
\BIBentryALTinterwordspacing
``{Carbon Trust: Route to Net Zero Standard},'' 2022. [Online]. Available:
  \url{https://www.carbontrust.com/}
\BIBentrySTDinterwordspacing

\bibitem{green-deal}
\BIBentryALTinterwordspacing
``{What is the impact of the war in Ukraine on Europe’s climate and energy
  policy?}'' 2022. [Online]. Available:
  \url{https://www.euronews.com/green/2022/03/24/what-is-the-impact-of-the-war-in-ukraine-on-europe-energy-policy}
\BIBentrySTDinterwordspacing

\bibitem{net-zero}
\BIBentryALTinterwordspacing
``{UK Net Zero Strategy: Build Back Greener},'' 2022. [Online]. Available:
  \url{https://www.gov.uk/government/publications/net-zero-strategy}
\BIBentrySTDinterwordspacing

\bibitem{lenovo-bigdata}
M.~Hodak, M.~Gorkovenko, and A.~Dholakia, ``Towards power efficiency in deep
  learning on data center hardware,'' in \emph{2019 IEEE International
  Conference on Big Data (Big Data)}, 2019.

\bibitem{Strubell2019}
E.~Strubell, A.~Ganesh, and A.~McCallum, ``Energy and policy considerations for
  deep learning in nlp,'' \emph{arXiv preprint arXiv:1906.02243}, 2019.

\bibitem{Ascierto2018}
\BIBentryALTinterwordspacing
R.~Ascierto, ``{Uptime Institute Global Data Center Survey 2018},'' 2018.
  [Online]. Available:
  \url{https://datacenter.com/wp-content/uploads/2018/11/2018-data-center-industry-survey.pdf}
\BIBentrySTDinterwordspacing

\bibitem{Schneider}
\BIBentryALTinterwordspacing
W.~Torell, ``{A data-driven model to forecast energy consumption at the
  Edge},'' 2021. [Online]. Available:
  \url{https://www.datacenterdynamics.com/en/opinions/a-data-driven-model-to-forecast-energy-consumption-at-the-edge/}
\BIBentrySTDinterwordspacing

\bibitem{Qiu2020}
X.~Qiu, T.~Parcollet, D.~J. Beutel, T.~Topal, A.~Mathur, and N.~D. Lane, ``Can
  federated learning save the planet?'' \emph{arXiv preprint arXiv:2010.06537},
  2020.

\bibitem{eea}
\BIBentryALTinterwordspacing
E.~E. Agency, ``{CO2 emission intensity},'' 2020. [Online]. Available:
  \url{https://www.eea.europa.eu/data-and-maps/daviz/co2-emission-intensity-9}
\BIBentrySTDinterwordspacing

\bibitem{tf-bench}
\BIBentryALTinterwordspacing
Tensorflow, ``{ML Benchmarks},'' 2022. [Online]. Available:
  \url{https://github.com/tensorflow/benchmarks}
\BIBentrySTDinterwordspacing

\bibitem{tsoc}
\BIBentryALTinterwordspacing
``{Daily Energy Production in Cyprus},'' 2022. [Online]. Available:
  \url{https://tsoc.org.cy/electrical-system/total-daily-system-generation-on-the-transmission-system/}
\BIBentrySTDinterwordspacing

\bibitem{GreenAI}
R.~Schwartz, J.~Dodge, N.~A. Smith, and O.~Etzioni, ``Green ai,'' \emph{Commun.
  ACM}, vol.~63, no.~12, 2020.

\bibitem{tf-model-zoo}
\BIBentryALTinterwordspacing
``{TensorFlow Lite Model Zoo},'' 2022. [Online]. Available:
  \url{https://www.tensorflow.org/lite/models}
\BIBentrySTDinterwordspacing

\bibitem{Symeonides2020}
M.~Symeonides, Z.~Georgiou, D.~Trihinas, G.~Pallis, and M.~D. Dikaiakos,
  ``Fogify: A fog computing emulation framework,'' in \emph{2020 IEEE/ACM
  Symposium on Edge Computing (SEC)}, 2020.

\bibitem{trihinas2016}
D.~Trihinas, G.~Pallis, and M.~D. Dikaiakos, ``Monitoring elastically adaptive
  multi-cloud services,'' \emph{IEEE Transactions on Cloud Computing}, vol.~6,
  no.~3, 2016.

\bibitem{Gao2014}
J.~Gao, ``Machine learning applications for data center optimization,'' 2014.

\bibitem{Patterson2021}
\BIBentryALTinterwordspacing
D.~A. Patterson, J.~Gonzalez, Q.~V. Le, C.~Liang, L.~Munguia, D.~Rothchild,
  D.~R. So, M.~Texier, and J.~Dean, ``Carbon emissions and large neural network
  training,'' \emph{CoRR}, vol. abs/2104.10350, 2021. [Online]. Available:
  \url{https://arxiv.org/abs/2104.10350}
\BIBentrySTDinterwordspacing

\bibitem{electricitymaps}
\BIBentryALTinterwordspacing
E.~Maps. The resource for 24/7 electricity co2 data. [Online]. Available:
  \url{https://electricitymaps.com/}
\BIBentrySTDinterwordspacing

\bibitem{GoogleCarbonAwareComputing}
A.~Radovanovic, R.~Koningstein, I.~Schneider, B.~Chen, A.~Duarte, B.~Roy,
  D.~Xiao, M.~Haridasan, P.~Hung, N.~Care, S.~Talukdar, E.~Mullen, K.~Smith,
  M.~Cottman, and W.~Cirne, ``Carbon-aware computing for datacenters,''
  \emph{IEEE Transactions on Power Systems}, 2022.

\bibitem{WorkloadShiftingCarbonEmissions}
P.~Wiesner, I.~Behnke, D.~Scheinert, K.~Gontarska, and L.~Thamsen, ``Let's wait
  awhile: How temporal workload shifting can reduce carbon emissions in the
  cloud,'' in \emph{Proceedings of the 22nd International Middleware
  Conference}, ser. Middleware '21.\hskip 1em plus 0.5em minus 0.4em\relax ACM,
  2021.

\bibitem{acun2022carbon}
B.~Acun, B.~Lee, F.~Kazhamiaka, K.~Maeng, M.~Chakkaravarthy, U.~Gupta,
  D.~Brooks, and C.-J. Wu, ``Carbon explorer: A holistic approach for designing
  carbon aware datacenters,'' \emph{arXiv preprint arXiv:2201.10036}, 2022.

\bibitem{Trihinas2017}
D.~Trihinas, G.~Pallis, and M.~Dikaiakos, ``{ADMin:} adaptive monitoring
  dissemination for the internet of things,'' in \emph{IEEE INFOCOM 2017},
  2017.

\bibitem{hasenburg2019mockfog}
J.~Hasenburg, M.~Grambow, E.~Gr{\"u}newald, S.~Huk, and D.~Bermbach, ``Mockfog:
  Emulating fog computing infrastructure in the cloud,'' in \emph{2019 IEEE
  International Conference on Fog Computing (ICFC)}.\hskip 1em plus 0.5em minus
  0.4em\relax IEEE, 2019.

\bibitem{beilharz2021towards}
J.~Beilharz, P.~Wiesner, A.~Boockmeyer, F.~Brokhausen, I.~Behnke, R.~Schmid,
  L.~Pirl, and L.~Thamsen, ``Towards a staging environment for the internet of
  things,'' in \emph{2021 IEEE International Conference on Pervasive Computing
  and Communications Workshops and other Affiliated Events (PerCom
  Workshops)}.\hskip 1em plus 0.5em minus 0.4em\relax IEEE, 2021.

\bibitem{nikolaidis2021iotier}
F.~Nikolaidis, M.~Marazakis, and A.~Bilas, ``Iotier: A virtual testbed to
  evaluate systems for iot environments,'' in \emph{2021 IEEE/ACM 21st
  International Symposium on Cluster, Cloud and Internet Computing
  (CCGrid)}.\hskip 1em plus 0.5em minus 0.4em\relax IEEE, 2021.

\bibitem{sonmez2018edgecloudsim}
C.~Sonmez, A.~Ozgovde, and C.~Ersoy, ``Edgecloudsim: An environment for
  performance evaluation of edge computing systems,'' \emph{Transactions on
  Emerging Telecommunications Technologies}, 2018.

\bibitem{zeng2017}
Y.~Zeng and R.~Zhang, ``Energy-efficient uav communication with trajectory
  optimization,'' \emph{IEEE Transactions on Wireless Communications}, vol.~16,
  no.~6, 2017.

\bibitem{gupta2017ifogsim}
H.~Gupta, A.~Vahid~Dastjerdi, S.~K. Ghosh, and R.~Buyya, ``ifogsim: A toolkit
  for modeling and simulation of resource management techniques in the internet
  of things, edge and fog computing environments,'' \emph{Software: Practice
  and Experience}, 2017.

\bibitem{mechalikh2019pureedgesim}
C.~Mechalikh, H.~Taktak, and F.~Moussa, ``Pureedgesim: A simulation toolkit for
  performance evaluation of cloud, fog, and pure edge computing environments,''
  in \emph{2019 International Conference on High Performance Computing \&
  Simulation (HPCS)}.\hskip 1em plus 0.5em minus 0.4em\relax IEEE, 2019.

\bibitem{wiesner2021leaf}
P.~Wiesner and L.~Thamsen, ``Leaf: Simulating large energy-aware fog computing
  environments,'' in \emph{2021 IEEE 5th International Conference on Fog and
  Edge Computing (ICFEC)}.\hskip 1em plus 0.5em minus 0.4em\relax IEEE, 2021.

\bibitem{TeadsCarbonEstimator}
\BIBentryALTinterwordspacing
T.~Engineering. Carbon footprint estimator for aws instances. [Online].
  Available:
  \url{https://engineering.teads.com/sustainability/carbon-footprint-estimator-for-aws-instances/}
\BIBentrySTDinterwordspacing

\bibitem{Lacoste2019}
A.~Lacoste, A.~Luccioni, V.~Schmidt, and T.~Dandres, ``Quantifying the carbon
  emissions of machine learning,'' \emph{arXiv preprint arXiv:1910.09700},
  2019.

\bibitem{anthony2020carbontracker}
L.~F.~W. Anthony, B.~Kanding, and R.~Selvan, ``Carbontracker: Tracking and
  predicting the carbon footprint of training deep learning models,''
  \emph{arXiv preprint arXiv:2007.03051}, 2020.

\end{thebibliography}

\end{document}